\documentclass[
superscriptaddress,
amsmath,amssymb,
longbibliography,
aps,
prl,
]{revtex4-1}
\usepackage{graphicx}
\usepackage{dcolumn}
\usepackage{bm}
\usepackage{color}
\usepackage{hyperref}
\usepackage{xcolor}
\usepackage{braket} 
\usepackage{cancel}
\allowdisplaybreaks

\newcommand{\bfac}[0]{\langle b \rangle}
\newcommand{\aprate}[0]{\Gamma^{\mathrm{ap}}}
\newcommand{\eprate}[0]{\Gamma^{\mathrm{ep}}}
\newcommand{\apeprate}[0]{\Gamma^{\mathrm{ap/ep}}}

\begin{document}


\title{Quantum interference between fundamentally different processes is enabled by shaped input wavefunctions}

\author{Jeremy Lim}
\affiliation{%
Science, Mathematics and Technology, Singapore University of Technology and Design, 8 Somapah Road, Singapore 487372, Singapore
}%

\author{Suraj Kumar}
\affiliation{
School of Electrical and Electronic Engineering, Nanyang Technological University, 50 Nanyang Avenue, Singapore 639798, Singapore
}

\author{Yee Sin Ang}
\affiliation{%
Science, Mathematics and Technology, Singapore University of Technology and Design, 8 Somapah Road, Singapore 487372, Singapore
}%

\author{Lay Kee Ang}
\affiliation{%
Science, Mathematics and Technology, Singapore University of Technology and Design, 8 Somapah Road, Singapore 487372, Singapore
}%

\author{Liang Jie Wong}
\email{liangjie.wong@ntu.edu.sg}
\affiliation{
School of Electrical and Electronic Engineering, Nanyang Technological University, 50 Nanyang Avenue, Singapore 639798, Singapore
}


\begin{abstract}
We present a general framework for quantum interference (QI) between multiple, fundamentally different processes. Our framework reveals the importance of shaped input wavefunctions in enabling QI, and predicts unprecedented interactions between free electrons, bound electrons, and photons: (i) the vanishing of the zero-loss peak by destructive QI when a shaped electron wavepacket couples to light, under conditions where the electron's zero-loss peak otherwise dominates; (ii) QI between free electron and atomic (bound electron) spontaneous emission processes, which can be significant even when the free electron and atom are far apart, breaking the common notion that electron and atom must be close by to significantly affect each other's processes. Our work shows that emerging quantum waveshaping techniques unlock the door to greater versatility in light-matter interactions and other quantum processes in general.
\\\\
Short title: A general framework for quantum interference\\\\
\end{abstract}

\pacs{Valid PACS appear here}
\maketitle
\section{Introduction} 
Interest in controlling quantum processes has led people to seek increasingly precise ways of manipulating the wavefunctions involved -- a process known as waveshaping. Through waveshaping, many unique wave patterns for particles including photons and electrons have been realized by introducing a well defined phase relation between different eigenstates. Shaped free electron wavepackets~\cite{Barkwick2009, Park2010, Kaminer2015a, Feist2015, Piazza2015, Priebe2017, vanacore2018, Liu2019, Wang2020, Harvey2020, Dahan2020, Liebtrau2021, GarciaDeAbajo2021, Vanacore2020, Yalunin2021, Feist2015, Zhao2021b, Morimoto2018, LJ2015, Priebe2017, Kozak2018, Kozak2018b, Lim2019, Harris2015, Kaminer2015, vanacore2018, Grillo2014, DiGiulio2019, Vanacore2019, Kfir2019, Kfir2020}, for instance, are useful as probes of light-matter excitations electron energy-loss spectroscopy (EELS) and its variants~\cite{VERBEECK2004207, GarciaDeAbajo2010, Verbeeck2010, VanTendeloo2012, EGOAVIL20141, Krehl2018, Polman2019, Liu2019, Liebtrau2021, GarciaDeAbajo2021, lopez2021},  cathodoluminescence (CL) microscopy~\cite{GarciaDeAbajo2010, DiGiulio2021, Liebtrau2021, GarciaDeAbajo2021} and photon-induced near-field electron microscopy (PINEM)~\cite{Barkwick2009, Park2010, Feist2015, Piazza2015, Priebe2017, vanacore2018, Liu2019, Wang2020, Harvey2020, Dahan2020, Liebtrau2021, GarciaDeAbajo2021}, etc.; and also as a means of tailoring light emission from free electrons~\cite{Guzzinati2017, Roques2018, Karnieli2021, Karnieli2021b, Gover2018, Pan2018, Pan2019, Gover2019, Faresab2015, Talebi2016, DiGiulio2021, Kfir2021, Wong2021}. Likewise, shaped bound electron (i.e., atomic states) form the basis of many fields including quantum metrology~\cite{Giovannetti2011, GLICKMANSHLOMI2013, Matthews2016, Sun2018, sanchez2021, GLICKMANSHLOMI2013}, quantum information technologies~\cite{Cirac1997, Boozer2007, Kimble2008, Ritter2012, HOFMANNMICHAEL2012, Stute2013, Northup2014, Heshami2016, Bechler2018}, quantum integrated circuits~\cite{CAROLAN2015, Kim2016, Harris2017, Uppu2020, Uppu2021}, and photon generation and manipulation~\cite{Kim2016, Michler2000, Faraon2008, Wu2009, Reinhard2012, Chu2017, Prasad2020, Stiesdal2021}. Many techniques exist to shape photons and quasi-particles like polaritons~\cite{Pursley2018, Morin2019, LIB2020, Uria2020, Kondakci2017, Bhaduri2020, Wong2020advsci, Xu2014, Machado2018, Estrecho2021}. Recently, shaped neutron wavefunctions -- especially twisted states -- have garnered interest as possible probes for nuclear structure and interactions~\cite{Larocque2018, Afanasev2019, Afanasev2021}, and for neutron interferometry and  optics~\cite{Clark2015,  Nsofini2016, Pushin2017, Cappelletti2018, Geerits2021}. 

Here, we show that shaped input wavefunctions are key to enabling quantum interference (QI) between processes that normally (i.e. for unshaped wavefunctions) do not affect each other significantly. It should be noted that specific examples of QI -- such as electromagnetically induced transparency~\cite{Harris1990, Boller1991}, QI between nuclear and electromagnetic processes~\cite{Schwinger1948, Palffy2007, Afanasev2019}, and QI between photons~\cite{Craddock2019} -- have been previously studied. However, a general framework for QI is lacking and the role of shaped input wavefunctions has not been elucidated. This has prevented the application of QI to the full range of existing quantum processes. We therefore present a general framework for QI between arbitrary types and numbers of quantum systems, where the presence of shaped input wavefunctions emerges as a fundamental requirement for QI. We also see that QI between more than two processes is possible, leading to a dominance of QI effects as the number of systems with shaped input wavefunctions increases. 

We use two examples to show that our framework can predict unprecedented phenomena in light-matter interaction. In the first example, we show that QI can eliminate the zero-loss peak of the output electron spectrum in free electron-light interactions at moderate coupling strengths. This is significant because the zero-loss peak is typically dominant in free electron gain/loss spectra in PINEM experiments under the conditions we consider. We also show that QI can dramatically enhance or suppress the other peaks in the free electron gain/loss spectrum.  Our results are achievable with parameters well within the capabilities of current PINEM setups.  In the second example, we show that QI can occur between free electron and atomic (a.k.a., bound electron) spontaneous emission (SE) processes. We choose this example for the following reasons: (i) There has been much excitement and progress in shaping electron wavepackets both spatially and temporally~\cite{Barkwick2009, Park2010, Kaminer2015a, Feist2015, Piazza2015, Priebe2017, vanacore2018, Liu2019, Wang2020, Harvey2020, Dahan2020, Liebtrau2021, GarciaDeAbajo2021, Vanacore2020, Yalunin2021, Feist2015, Zhao2021b, Morimoto2018, LJ2015, Priebe2017, Kozak2018, Kozak2018b, Lim2019, Harris2015, Kaminer2015, vanacore2018, Grillo2014, DiGiulio2019, Vanacore2019, Kfir2019, Kfir2020}, but the QI this enables with bound electron processes, such as atomic SE, has never been explored; (ii) Free electron and  bound electron SE processes are not expected to significantly influence each other unless the free electron and bound electron are very close by, due to the near-field nature of Coulomb interactions~\cite{Gover2020, Ruimy2021, Zhao2021a, Zhang2021, Zhang2021b, Ratzel2021}. On the contrary, we find that owing to QI, shaped free electrons and shaped bound electrons can affect each  other even when both systems are physically far apart. Our results show that maximum SE enhancement or suppression can be achieved over a wide range of free electron kinetic energies (e.g., 100 eV to 1 MeV) and emission frequencies (e.g., optical to terahertz). Our findings fill an important gap in the understanding  of QI. Our work also motivates the development of shaping techniques for a wider variety of quantum systems to leverage the full potential of QI for on-demand tailoring of quantum processes in light-matter interactions and beyond.

\section{Results}

\textbf{General framework for QI} Consider a collection of $N$ distinct systems, e.g., free electrons, bound electrons, photons, atomic nuclei, neutrons, and any other fundamental and quasi-particles. We denote the eigenstates of the $j$th system as $\ket{\alpha_{j}}$ with corresponding eigenvalues $\alpha_{j}$. We consider an initial state of the form ${\ket{\text{initial}} = \bigotimes_{j=1}^{N}(\sum_{\alpha_{j}}C_{\alpha_{j}}\ket{\alpha_{j}})}$, where $|C_{\alpha_{j}}|^{2}$ is the probability of finding the $j$th system in $\ket{\alpha_{j}}$. The probability that $\ket{\text{initial}}$ scatters into a final state $\ket{\text{final}} =\ket{\beta_{1},...,\beta_{N}}$ after an arbitrary interaction, described by the scattering operator $\hat{\mathcal{S}}$, is $P_{\text{final}} = |\braket{\text{final}|\hat{\mathcal{S}}|\text{initial}}|^{2}$. Expanding $\ket{\text{initial}}$ in full and defining the coherence and population of the $j$th system as $\rho_{\alpha_{j}\alpha_{j}'} \equiv C_{\alpha_{j}}\bar{C}_{\alpha_{j}'}$ (where $\alpha_{j}\neq \alpha_{j}'$) and $p_{\alpha_{j}} \equiv \rho_{\alpha_{j}\alpha_{j}}$ (overbars denote complex conjugates) respectively, we express $P_{\text{final}}$ as
\begin{widetext}
\begin{equation}
\begin{split}
P_{\text{final}} &= \overbrace{\sum_{\alpha_{1},...,\alpha_{N}} p_{\alpha_{1}}...p_{\alpha_{N}}\left|\mathcal{S}^{\text{final}}_{\alpha_{1},...,\alpha_{N}}\right|^{2}}^{\text{without QI}} +\overbrace{\bigg{[}{N\choose 1}\text{ terms of}\sum_{\substack{\text{all except}\\\alpha_{j}}}\sum_{\substack{\alpha_{j}\neq\alpha_{j}'}}\rho_{\alpha_{j}\alpha_{j}'}\frac{p_{\alpha_{1}}...p_{\alpha_{N}}}{p_{\alpha_{j}}}\mathcal{S}^{\text{final}}_{...,\alpha_{j},...}\bar{\mathcal{S}}^{\text{final}}_{...,\alpha_{j}',...} \Bigg{]}}^{\text{1-process QI}} \\
& +\underbrace{\bigg{[}{N\choose 2}\text{ terms of}\sum_{\substack{\text{all except}\\\alpha_{j},\alpha_{k}}}\sum_{\substack{\alpha_{j}\neq\alpha_{j}' \\ \alpha_{k}\neq\alpha_{k}'}}\rho_{\alpha_{j}\alpha_{j}'}\rho_{\alpha_{k}\alpha_{k}'}\frac{p_{\alpha_{1}}...p_{\alpha_{N}}}{p_{\alpha_{j}}p_{\alpha_{k}}}\mathcal{S}^{\text{final}}_{...,\alpha_{j},\alpha_{k},...}\bar{\mathcal{S}}^{\text{final}}_{...,\alpha_{j}',\alpha_{k}',...} \Bigg{]}}_{\text{2-process QI}} \\
&  + \underbrace{\Bigg{[}{N\choose 3}\text{ terms of}\sum_{\substack{\text{all except}\\\alpha_{j},\alpha_{k},\alpha_{l}}}\sum_{\substack{\alpha_{j}\neq\alpha_{j}' \\ \alpha_{k}\neq\alpha_{k}' \\ \alpha_{l}\neq\alpha_{l}'}}\rho_{\alpha_{j}\alpha_{j}'}\rho_{\alpha_{k}\alpha_{k}'}\rho_{\alpha_{l}\alpha_{l}'}\frac{p_{\alpha_{1}}...p_{\alpha_{N}}}{p_{\alpha_{j}}p_{\alpha_{k}}p_{\alpha_{l}}}\mathcal{S}^{\text{final}}_{...,\alpha_{j},\alpha_{k},\alpha_{l},...}\bar{\mathcal{S}}^{\text{final}}_{...,\alpha_{j}',\alpha_{k}',\alpha_{l}',...}\Bigg{]}}_{\text{3-process QI}} \\
& + \quad\dots\quad +\underbrace{\Bigg{[}\sum_{\alpha_{1}\neq\alpha_{1}',...,\alpha_{N}\neq\alpha_{N}'}\rho_{\alpha_{1}\alpha_{1}'}\dots\rho_{\alpha_{N}\alpha_{N}'}\mathcal{S}^{\text{final}}_{\alpha_{1},...,\alpha_{N}}\bar{\mathcal{S}}^{\text{final}}_{\alpha_{1}',...,\alpha_{N}'}\Bigg{]}}_{\text{$N$-process QI}},
\end{split}
\label{eqn_Pfinal_full_expansion}
\end{equation}
\end{widetext}
where $\mathcal{S}^{\text{final}}_{\alpha_{1},...,\alpha_{N}} \equiv \braket{\text{final}|\hat{\mathcal{S}}|\alpha_{1},..,\alpha_{N}}$ is the S-matrix element.  The total scattering probability into states that share the same final values of quantum numbers $\beta_{m},...,\beta_{n}$ is $P_{\beta_{m},...,\beta_{n}} = \sum_{\text{all except}\beta_{m},...,\beta_{n}}P_{\text{final}}$. The first term of Eq.~(\ref{eqn_Pfinal_full_expansion}) is the total probability in the absence of QI. Note that the absence of QI corresponds to the scenario where multiple eigenstates exist, but they are related by random phases, resulting in QI disappearing upon statistical averaging. The terms in the $R$th square parentheses, where $R\in\mathbb{Z}^{+}$, contain the ${N\choose R}$ possible QI terms that can arise between $R$ of the $N$ systems. Crucially, our framework reveals the importance of shaped input wavefunctions as a prerequisite for (and a means to tailor) QI: the initial wavefunctions of the systems participating in QI must be a superposition of eigenstates with well-defined phase relations between them, such that the coherences of the systems involved in the QI -- and hence the relevant QI terms in Eq.~(\ref{eqn_Pfinal_full_expansion}) -- are non-zero upon statistical averaging. 

\begin{figure*}[ht!]
\centering
\includegraphics[width=0.98\textwidth]{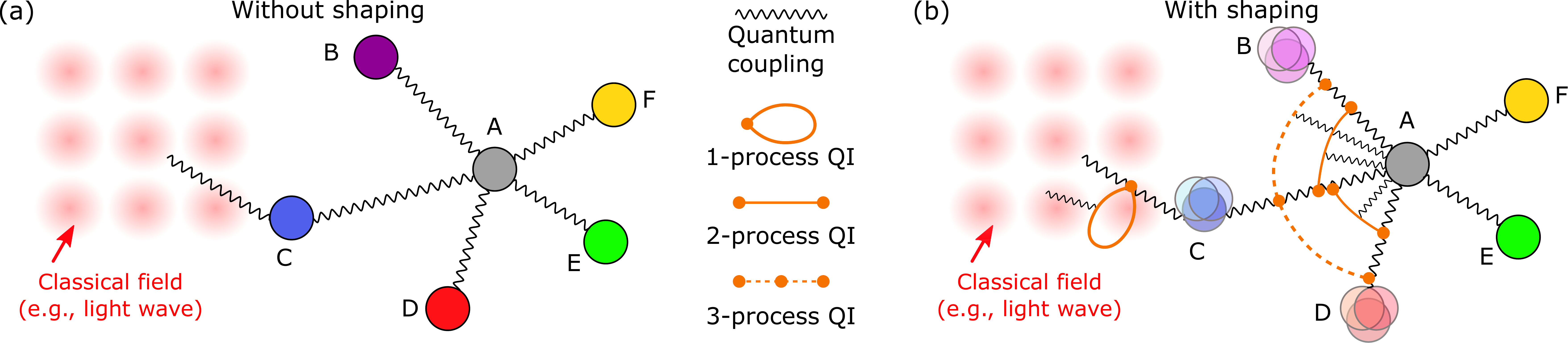}
\caption{A general framework for quantum interference (QI). QI is enabled by shaped input quantum wavefunctions, providing a means to tailor quantum processes via quantum waveshaping.  To illustrate this, we consider the example of 5 systems (B-F) coupled to system A. (a) When systems B to F have unshaped (i.e., either single-input state or a superposition of eigenstates with random phase relations, represented by solid circles) input wavefunctions, QI is absent and there are only direct coupling pathways between B-F and A. As a result the total coupling between system A and the other systems is simply the sum of the individual direct coupling terms considered in isolation. (b) When some of the systems (B-D here) have shaped input wavefunctions (i.e., superposition of eigenstates with fixed phase relations, denoted by overlapping transluscent circles), QI occurs between the processes associated with the shaped wavefunctions, resulting in additional coupling pathways. Notably, QI can involve more than two processes in general, resulting in a dominance in the number of QI terms as the number of systems with shaped wavefunctions increases. Note that 1-process QI can occur for instance, when the scattering events involve classical fields (e.g., self-loop in coupling between system C and classical field).}
\label{fig_1}
\end{figure*}  

Figure \ref{fig_1} illustrates this by considering a system A coupled in a pairwise manner to systems B to F (note in general that Eq.~(\ref{eqn_Pfinal_full_expansion}) is not limited to pairwise coupling). We see that shaping the input wavefunctions of systems B to D results in 2-process and 3-process QI between the processes associated with these shaped wavefunctions, providing additional coupling pathways to couple to A.  1-process QI can arise, for instance, when a system C with its input wavefunction shaped couples to classical fields (e.g., a light wave). 
\centerline{}

\begin{figure*}[ht!]
\centering
\includegraphics[width=0.95\textwidth]{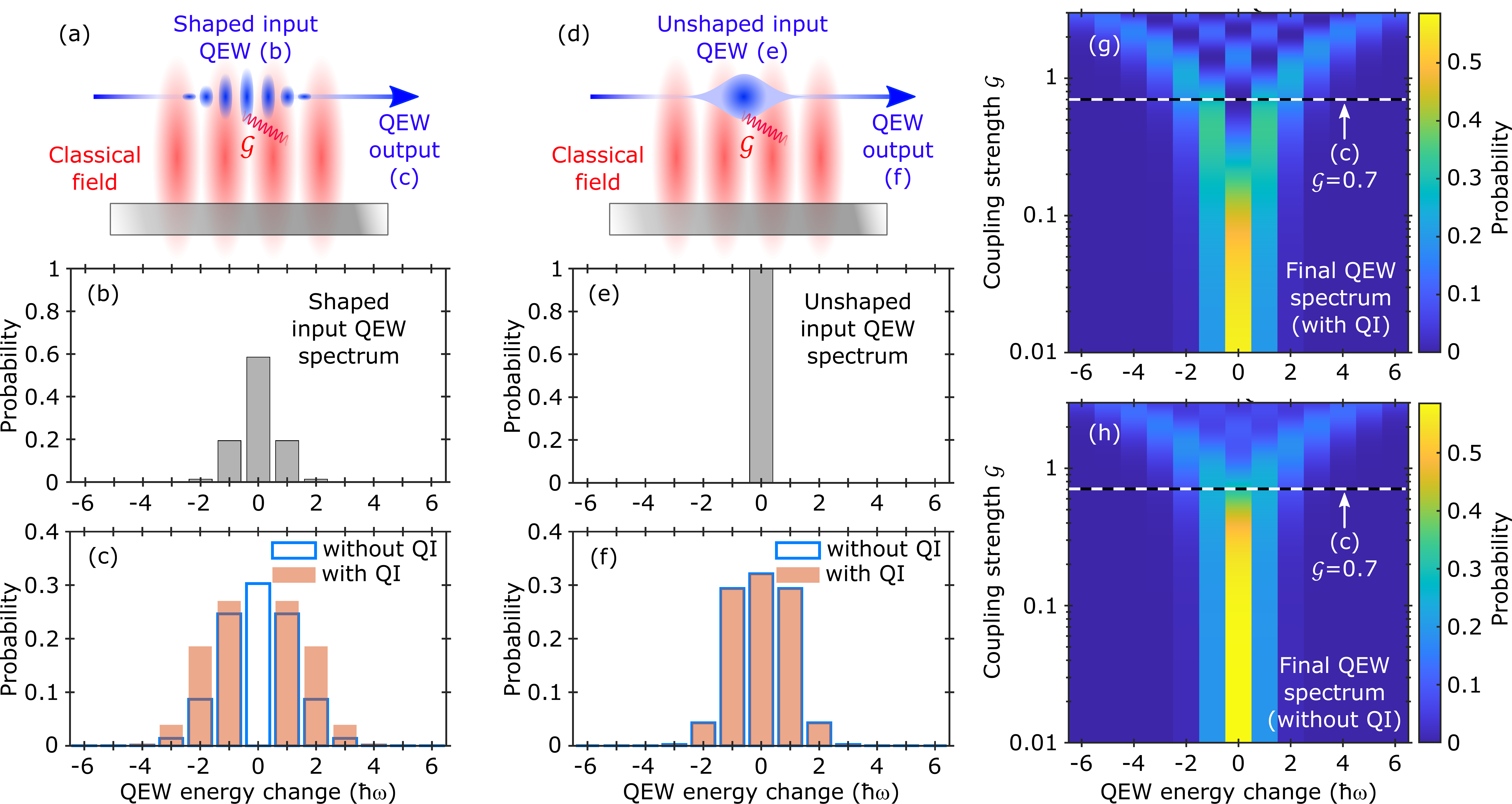}
\caption{Quantum interference (QI) in free electron-light interactions at moderate coupling strengths, resulting in elimination of the zero-loss peak (ZLP) by destructive QI, and enhancement of satellite peaks by constructive QI. (a) An incoming shaped free electron (modeled as a quantum electron wavepacket (QEW)), with input spectrum shown in (b), scatters off a classical light field with a dimensionless coupling strength $\mathcal{G} = 0.7$, resulting in an output electron spectrum where the ZLP completely vanishes. If QI contributions are neglected (unfilled bars with blue outlines in (c)), which occurs, for instance, when there is no fixed phase relation between the input eigenstates, the ZLP dominates. (d)~In contrast, for an unshaped incoming QEW (input spectrum shown in (e)), the output QEW spectrum in the presence and absence of QI coincide (as shown in (f)), which implies that QI contributions vanish for unshaped input QEWs. (g) and (h) compare the output QEW spectrum in the presence and absence of QI, respectively, as a function of $\mathcal{G}$. The contribution of QI is already substantial even for weaker interactions of about $\mathcal{G}\sim 0.1$. As shown in (h), the complete suppression of gain/loss peaks away from the ZLP can be achieved for coupling strengths $\mathcal{G}\gtrsim 1$.}
\label{fig_2}
\end{figure*}

\textbf{Elimination of zero-loss peak by QI in free electron-light interactions} When a shaped incoming free electron, i.e., a quantum electron wavepacket (QEW), is scattered by a classical light wave at moderate coupling strengths (Figs.~\ref{fig_2}(a),(b)), the output energy gain/loss spectrum (orange bars in Fig.~\ref{fig_2}(c)) shows a complete suppression of the zero-loss peak (ZLP) -- this is a direct consequence of QI. If QI is neglected the ZLP remains dominant in the output spectrum (blue-outlined unfilled bars in Fig.~\ref{fig_2}(c)). Note that neglecting QI corresponds to the physical scenario where a random phase relation exists between the input electron eigenstates, which results in the disappearance of QI effects upon statistical averaging. In addition to the complete suppression of the ZLP, QI also enhances the gain/loss peaks away from the ZLP. In contrast, for an unshaped incoming QEW (Figs.~\ref{fig_2}(d),(e)), there are no QI contributions and the ZLP dominates in the output spectrum (Fig.~\ref{fig_2}(f)). We model the wavefunction of the incoming QEW with the general form $\ket{\mathrm{initial}} = \sum_{n}C_{n}\ket{n}$, where $C_{n} = e^{i\phi_{\mathrm{mod}}}J_{n}(2|\mathcal{G}_{\mathrm{mod}}|)$ is the initial complex amplitude of the $n$th energy gain/loss peak of the incoming QEW~\cite{Feist2015},  $\phi_{\mathrm{mod}}$ is the modulation phase ($\phi_{\mathrm{mod}} = 0$), $J_{n}()$ is the Bessel function of the first kind, and $\mathcal{G}_{\mathrm{mod}}$ is the dimensionless coupling strength of the shaping stage ($\mathcal{G}_{\mathrm{mod}} = 0.5$ in (a)-(c),(g),(h) and $\mathcal{G}_{\text{mod}} = 0$ in (d)-(f)). The exact QEW-light interaction is described by the S-matrix $\hat{\mathcal{S}}\equiv \exp(\mathcal{G}^{*}\hat{b} - \mathcal{G}\hat{b}^{\dagger})$~\cite{Feist2015}, where $\mathcal{G}$ is the dimensionless coupling strength between the incoming QEW and the classical light field. Here, $\hat{b}$ ($\hat{b}^{\dagger}$) decrements (increments) each QEW eigenstate by a unit photon energy. We obtain the $N$th energy gain/loss peak of the final output QEW probability spectrum (i.e., the probability that $\ket{\text{initial}}$ scatters into a final state $\ket{N}$) as
\begin{equation}
P_{N} = \underbrace{\sum_{n}|C_{n}|^{2}|\braket{N|\hat{\mathcal{S}}|n}|^{2}}_{\text{without QI}} +\underbrace{\sum_{m\neq n}C_{m}C_{n}^{*}\braket{N|\hat{\mathcal{S}}|m}\braket{N|\hat{\mathcal{S}}|n}^{*}}_{\text{1-process QI}},\quad m,n,N\in\mathbb{Z},
\label{eqn_single_QEW}
\end{equation}
which is the sum of the first and second terms in Eq.~(\ref{eqn_Pfinal_full_expansion}) with only a single system (the QEW) as input. In Figs.~\ref{fig_2}(a)-(f), we consider a coupling strength of $\mathcal{G} = 0.7$, which is well within the reach of existing PINEM setups, where coupling strengths on the order of $\mathcal{G}\sim 100$ has been demonstrated~\cite{Dahan2020}. For smaller coupling strengths of $\mathcal{G}\gtrsim 0.1$, the ZLP appears but remains suppressed by destructive QI, as we show in Figs.~\ref{fig_2}(g) and \ref{fig_2}(h). Additionally, for $\mathcal{G} \gtrsim 1$, complete suppression of the gain/loss peaks away from the ZLP can also occur due to QI. We find that our results still hold for incoming QEWs of different shapes (Supplemental Material (SM) Section I).

\begin{figure*}[ht!]
\centering
\includegraphics[width=1\textwidth]{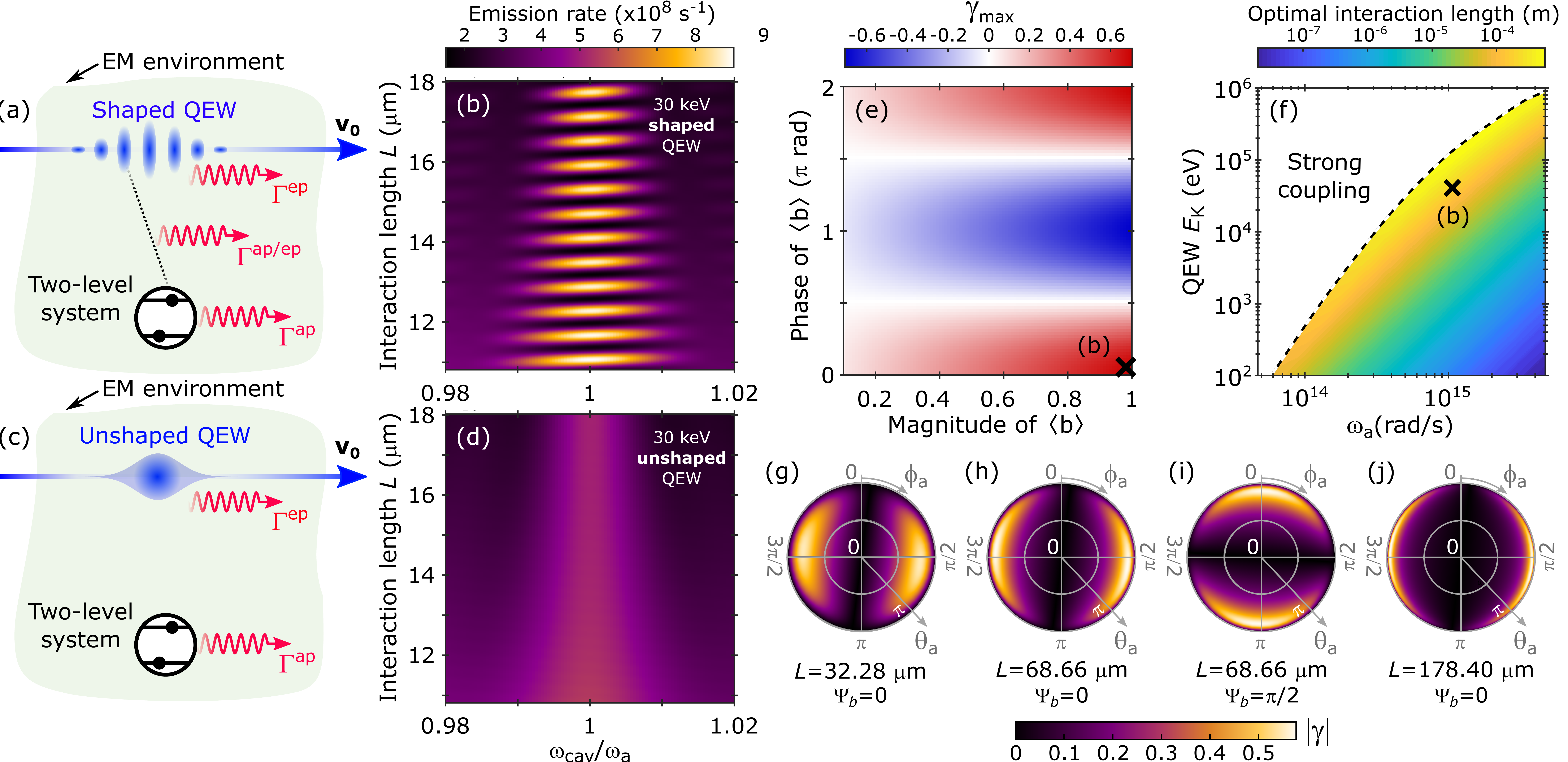}
\caption{Quantum interference (QI) provides a means to tailor spontaneous emission from free electrons and bound electrons via quantum waveshaping. (a) Shaped quantum electron wavepackets (QEWs) and bound electrons (atomic two-level system) within an electromagnetic (EM) environment separately emit photons at rates of $\eprate$ and $\aprate$ respectively. QI between these spontaneous emission (SE) processes (dotted line) results in a third emission process at rate $\apeprate$, which can enhance or suppress the total SE rate as a function of interaction length $L$ (as shown in (b)) by more than $70\%$. (c) In contrast, for unshaped incoming QEWs, QI is absent (as shown in (d)) and the total SE rate is $\aprate+\eprate$. QI can be tailored using the QEW and bound electron shapes, which are determined by bunching factor $\bfac = |\bfac|e^{i\Psi_{b}}$ and coherence $\rho_{eg}^{a} = |\rho_{eg}^{a}|e^{i\phi_{a}}$ respectively. Defining the figure of merit $\gamma= \apeprate/(\aprate + \eprate)$, which is a measure of the QI contribution, we see from (e) that larger $|\gamma_{\mathrm{max}}| = \mathrm{max}(|\gamma|)$ is achieved for larger $|\bfac|$, and that QI can be tuned to enhance or suppress the SE rate by controlling the phase of the bunching factor $\Psi_{b}$. (f) shows the dependence of the optimal length $L_{\mathrm{opt}}$ -- at which $\gamma_{max}$ is achieved -- as a function of QEW kinetic energy $E_{\mathrm{K}}$ and bound electron emission frequency $\omega_{a}$. The polar plots in (g)-(j) show the value of $|\gamma|$ on the Bloch sphere representing the initial shape of the bound electron system at various values of $L$, and (h),(i) shows that the profile of $|\gamma|$ on the initial Bloch sphere can be manipulated by varying $\Psi_{b}$. Unless otherwise stated, we consider a 30 keV shaped QEW with $\bfac = 0.99$. Our two-level, bound electron system is a Sn-N vacancy with $\omega_{a} \approx 3\times10^{15}$ rad/s and transition dipole moment of $|\mathbf{d}| = 4.33\times10^{-29}$ Cm aligned parallel to the field. For (g)-(j), the azimuthal angle $\phi_{a}$ is the phase of the bound electron coherence and $\theta_{a}$ is related to the excited state population through $\cos^{2}(\theta_{a}/2)$, and we use a 30 keV shaped QEW with $\bfac = 0.58$.}
\label{fig_3}
\end{figure*}

\textbf{QI between free electron and bound electron spontaneous emission processes} We now apply our framework to study QI between free electron and bound electron SE processes in a cavity (or any electromagnetic environment in general). A shaped incoming QEW (Fig.~\ref{fig_3}(a)) can induce a QI SE contribution $\apeprate$ which enhances or suppresses the sum of the individual SE processes $\aprate + \eprate$ by more than $70\%$, depending on interaction length $L$ and cavity angular frequency $\omega_{\text{cav}}$ (Fig.~\ref{fig_3}(b)). Here $\aprate$ and $\eprate$ denote the bound electron and free electron SE rates respectively.  In contrast, there is no QI for an unshaped (Gaussian) incoming QEW (Figs.~\ref{fig_3}(c),(d)). As expected, we find that the SE rates peak sharply at resonance ($\omega_{\mathrm{cav}} = \omega_{a}$).  We consider the resonant case for the rest of this example. The incoming QEW of velocity $\mathbf{v} = v_{0}\mathbf{\hat{z}}$ has a central kinetic energy of $E_{\text{K}} = 30$ keV. We treat the bound electron system as a two-level atomic system by considering a tin-vacancy (SnV) center~\cite{Trusheim2020} of emission frequency $\omega_{a} \approx 3\times 10^{15}$ rad/s and dipole moment $\mathbf{d} = \mathbf{\hat{z}}4.33\times 10^{-29}$ Cm (aligned parallel to the field).

Importantly, the shapes of the QEW and bound electron can be used to tailor $\apeprate$. We define the figure of merit $\gamma = \apeprate/(\aprate + \eprate)$ as a measure of QI's relative contribution to the SE rate. Figure~\ref{fig_3}(e) shows how the incoming QEW shape -- determined by the bunching factor $\bfac = |\bfac|e^{i\Psi_{b}}$ -- affects $\gamma_{\mathrm{max}}$, which is the maximum possible $\gamma$ across all $L$. For the case shown in Fig.~\ref{fig_3}(d) (black cross in Fig.~\ref{fig_3}(e)), we considered $\bfac = 0.99$, which has  recently been shown to be feasible~\cite{Yalunin2021}.  Even for a more modest bunching factor of $\bfac\approx 0.58$, which is attainable using PINEM (see SM Section II and \cite{Yalunin2021}), for instance, $|\gamma_{\mathrm{max}}| \gtrsim 0.4$ can still be achieved. Varying the phase $\Psi_{b}$ (e.g., using phase of the modulating field) also allows us to control the amount by which the overall spontaneous emission rate is enhanced or suppressed.  Figure.~\ref{fig_3}(f) shows that the optimal interaction length $L_{\mathrm{opt}}$ needed to achieve $|\gamma_{\mathrm{max}}|\approx 0.707$, which is on the order of $\sim10$ nm to $\sim100$ $\mu$m,  falls in the range of experimentally realizable optical~\cite{Vahala2003, Vaidya2018, Davis2019, Muniz2020} and terahertz~\cite{Todorov2009, Derntl2017, Lu2018, Liu2020, Messelot2020} waveguide/cavity dimensions. Furthermore, the required QEW energies, which range from $\sim$100 eV to $\sim1$ MeV, are achievable using lab-scale electron sources. This implies that it should be already feasible to perform experiments to observe QI in SE from superconducting qubits~\cite{Kjaergaard2020} and quantum dots~\cite{Patel2010, Flagg2010, Loredo2016, Somaschi1026, Ding2016, Wei2014, Aharonovich2016, Gao2015, Veldhorst2015}, which radiate in the terahertz and optical regimes (SM Section III). An approximate analytical expression for $L_{\mathrm{opt}}$ is presented in SM Section IV. 

The QI contribution can also be controlled via the bound electron shape, which is determined by the coherence $\rho_{eg}^{a} = |\rho_{eg}^{a}|e^{i\phi_{a}}$ between its excited and ground states. We show this in Figs.~\ref{fig_3}(g)-(j), which depict $|\gamma|$ as a function of the initial bound electron Bloch sphere at various values of $L$. Here, $\theta_{a}$ (radial coordinate) and $\phi_{a}$ (angular coordinate) are related to the excited state population $\rho_{ee}^{a}$ and coherence $\rho^{a}_{eg}$ through $\rho_{ee}^{a} = \cos^{2}(\theta_{a}/2)$ and $\rho_{eg}^{a} = (e^{i\phi_{a}}/2)\sin\theta_{a}$ respectively. The bunching factor phase $\Psi_{b}$ can be used to azimuthally rotate the profile of $\gamma$ on the bound electron Bloch sphere, as seen in Figs~\ref{fig_3}(h),(i).

For Fig.~\ref{fig_3}, we model the QI contribution $\apeprate$ by considering a QEW of velocity $\mathbf{v} = v_{0}\mathbf{\hat{z}}$,  modulated at frequency $\omega_{\mathrm{mod}}$ and bunching factor $\bfac = |\bfac|e^{i\Psi_{b}}$ passing through an electromagnetic (EM) environment (e.g., cavity, waveguide) of length $L$ (also the interaction length) containing a bound electron system of coherence $\rho_{eg}^{a} = |\rho_{eg}^{a}|e^{i\phi_{a}}$. To first order in perturbation theory (i.e., weak coupling regime), we find that the single-photon SE rate arising from the QI between the free electron and bound electron SE processes is (SM Section V)
\begin{equation}
\begin{split}
\apeprate =& \frac{\tau}{\hbar}\frac{ev_{0}\omega_{a}|\mathbf{d}|}{\epsilon_{0}V\omega_{\mathrm{cav}}}|\rho^{a}_{eg}||\bfac|\cos(\xi)\mathrm{sinc}\bigg{[}\frac{(\omega_{\mathrm{cav}} - \omega_{a})\tau}{2}\bigg{]} \mathrm{sinc}\bigg{[}\frac{(\beta_{0}\omega_{\mathrm{cav}} - \omega_{\mathrm{mod}})\tau}{2}\bigg{]}\mathrm{sinc}\bigg{[}\frac{(\omega_{\mathrm{cav}} - \omega_{\mathrm{mod}})\tau}{2}\bigg{]},
\end{split}
\tag{2}
\label{eqn_gamma_elal}
\end{equation}
where the bound electron is located at $\mathbf{r} = (0,0,z_{a})$. The EM environment supports a single dominant longitundinal field mode of angular frequency $\omega_{\mathrm{cav}}$ and wavevector $\mathbf{q} = (0,0,\omega_{\mathrm{cav}}/c)$, where $c$ is the free-space speed of light. Such a mode is realizable, for instance, using a racetrack waveguide~\cite{Kfir2019}.  In Eq.~(\ref{eqn_gamma_elal}), $\hbar$ is the reduced Planck constant, $\epsilon_{0}$ is the free-space permitivitty, $e>0$ is the elementary charge, $\tau=L/v_{0}$ is the interaction duration, $\mathbf{d}$ is the bound electron transition dipole moment, $V$ is the mode volume, $m_{e}$ is the electron rest mass, $\beta_{0} = v_{0}/c$ is the normalized free electron velocity, $\xi = \phi_{a} - (\omega_{\mathrm{cav}}z_{a}/c)  - \pi/2 + \Psi_{b}$, and $\phi_{a}$ is the  bound electron coherence phase. The total SE rate is simply $\aprate+\eprate+\apeprate$. Unless otherwise stated, the initial excited and ground state populations are equal, corresponding to coherence magnitude $|\rho_{eg}^{a}| = 1/2$. We set $\phi_{a} - (\omega_{\mathrm{cav}}z_{a}/c) = \pi/2$,  which maximizes the contribution of $\apeprate$.  Note that $\aprate+\eprate$ and $\apeprate$ are derived from the first and third terms of Eq.~(\ref{eqn_Pfinal_full_expansion}) respectively when 3 systems (QEW, bound electron, and photon) are considered and the final state is summed over all possible output states containing 1 photon.  The second term in Eq.~(\ref{eqn_Pfinal_full_expansion}) (1-process QI) vanishes. Importantly, we see from Eq.~(\ref{eqn_gamma_elal}) that if either the free electron state or bound electron state (or both) is unshaped  (corresponding to $\bfac = 0$ and $\rho_{eg}^{a} = 0$ respectively),  $\apeprate = 0$ and the QI contribution vanishes, which we expect.

In our specific example of QI between free electron SE and bound electron SE, we find that QI affects the total SE rate substantially even when the free electron and bound electron systems are physically far apart. This is in sharp contrast to the Coulomb interaction between the atomic system and the QEW, which relies on the proximity between the atom and the QEW, and which has been leveraged in free-electron-bound-electron-resonant interaction (FEBERI)~\cite{Gover2020, Ruimy2021, Zhao2021a, Zhang2021, Zhang2021b, Ratzel2021} to encode information on bound electron coherence and dephasing in electron spectra.  In this regard, our work could provide a complementary route towards free-electron quantum metrology without the requirement of the bound electron system and QEW being physically near each other.

\section{Discussion} 
In essence, the far-reaching implications of our general QI framework are as follows: (i) Fundamentally distinct quantum processes can be made to affect each other through QI; (ii) In the presence of multiple shaped wavefunctions, multiple types of QI can arise, which can lead to the dominance in the number of QI terms in the overall output rate; (iii) QI can exist not only between quantum systems, but also between quantum systems and classical fields. These QI-driven effects are enabled by shaped wavefunctions -- a fundamental requirement revealed by our general framework, which provides the connection between shaped input wavefunctions and QI, showing that one implies the another. This motivates the development of innovative shaping techniques for fundamental particles and other quantum systems. Neutrons, for instance, with their ability to couple to all four fundamental forces, hypothetical particles (e.g., dark matter, axions) and interactions (e.g., modified gravity)~\cite{Sponar2021}, are promising testbeds for the foundations of cosmology and quantum mechanics. Ongoing efforts to shape the neutron~\cite{Clark2015, Cappelletti2018} imply that it may soon be possible to observe QI between neutron-driven processes and other types of processes. This could, for instance, provide additional degrees of freedom through which these exotic interactions and particles can be probed.

While we have only considered unentangled input quantum states, by using a more general expression for the input state $\ket{\text{initial}} = \sum_{\alpha_{1},...,\alpha_{N}}C_{\alpha_{1},...,\alpha_{N}}\ket{\alpha_{1}}\otimes ...\otimes\ket{\alpha_{N}}$, where $C_{\alpha_{1},...,\alpha_{N}} \neq C_{\alpha_{1}}...C_{\alpha_{N}}$ in general, our framework can be readily extended to study the effects of entangled input states, opening up a rich field of exploration. Our findings also suggest exciting prospects for applying QEWs that go beyond controlling photon emission, such as interference between free electron-photon and free electron-bound electron interactions for manipulation of free electron wavepackets and photon statistics, and interference between free electron-bound electron and bound electron-photon interactions for manipulation of bound electron population and coherence.

The general framework for QI we present also subsumes phenomena like electromagnetically induced transparency~\cite{Harris1990, Boller1991}, where destructive QI between transition amplitudes in a three-level system renders the system transparent in a spectral window, as well as weakly coupled free electron-photon interactions~\cite{Pan2019,Pan2019b}, where QI is analyzed as mixed-order terms arising from the interference between orders of a perturbative series expansion. Our framework goes far beyond the prediction of these phenomena, as we show through two unprecedented physical interactions: the vanishing of zero-loss peak in electron-light interactions for moderate coupling strengths, and QI between free electrons and bound electron SE processes. Unlike any other existing framework, our framework shows that shaped wavefunctions are essential for QI, and can enable QI between fundamentally different processes.

In summary, we present a general framework that describes QI between arbitrary types and numbers of quantum systems, revealing that shaped input wavefunctions are necessary in enabling QI. Intriguingly, we find that QI between more than two processes is possible, leading to a dominance in the number of QI terms as the number of shaped input increases. We apply our general QI framework to two examples and predict unprecedented phenomena in light-matter interactions. In the first example, we show using experimentally realistic parameters that QI can eliminate the zero-loss peak of the output free electron spectrum in free electron-light interactions at moderate coupling strengths -- a significant result since the zero-loss peak is typically dominant in the output electron spectra of PINEM. We also show that QI can dramatically enhance or suppress the satellite peaks in the output free electron spectrum. In the second example, we show that QI can occur between shaped free electron and bound electron SE processes even when both systems are physically distant and not able to interact via the Coulomb force.  We find that the total SE rate can be enhanced or suppressed by up to $70\%$ relative to the sum of isolated free electron and bound electron SE rates as a direct consequence of QI.  Our findings fill an important gap in the understanding of waveshaping as a prerequisite of QI, and introduce unexplored methods of tailoring and optimizing quantum interactions. Coupled with growing interest in shaping a wide range of quantum systems, including free electrons and neutrons, our work unlocks the possibility of QI for on-demand of tailoring light-matter interactions and beyond.


%
%
%
%

\begin{acknowledgements}
\textbf{Acknowledgements}  We thank I. Kaminer and N. Rivera for insightful discussions. L.J.W. acknowledges the support of the National Research Foundation (Project ID NRF2020-NRF-ISF004-3525) and the Nanyang Assistant Professorship Start-up Grant.  J.L. and L.K.A. acknowledge funding from A*STAR AME IRG (Project ID A2083c0057), MOE PhD Research Scholarship, and USA Office of Naval Research (Global) grant (Project ID N62909-19-1-2047). Y.S.A. acknowledges funding from SUTD Startup Research Grant (Project ID SRT3CI21163).
\end{acknowledgements}



%

\end{document}